\documentclass[amsmath,amssymb,aps,pra,floatfix,
twocolumn,superscriptaddress]{revtex4-1}

\usepackage{color}
\usepackage{graphicx}
\usepackage{physics}
\usepackage{blindtext}
\usepackage[normalem]{ulem}

\newcommand{\ba}{\begin{eqnarray}}
\newcommand{\ea}{\end{eqnarray}}

\begin{document}

\title{Stretching potential engineering}

\author{Stefano Cusumano}
\affiliation{NEST, Scuola Normale Superiore and Istituto Nanoscienze-CNR, I-56127 Pisa, Italy}
\email{stefano.cusumano@sns.it}

\author{Antonella De Pasquale}
\affiliation{Dipartimento di Fisica e Astronomia, Universit\'a di Firenze, I-50019, Sesto Fiorentino (FI), Italy}
\affiliation{INFN Sezione di Firenze, via G.Sansone 1, I-50019 Sesto Fiorentino (FI), Italy}
\affiliation{NEST, Scuola Normale Superiore and Istituto Nanoscienze-CNR, I-56127 Pisa, Italy}
\email{antonella.depasquale@sns.it}

\author{Giuseppe Carlo La Rocca}
\affiliation{Scuola Normale Superiore and CNISM, 56126 Pisa, Italy}

\author{Vittorio Giovannetti}
\affiliation{NEST, Scuola Normale Superiore and Istituto Nanoscienze-CNR, I-56127 Pisa, Italy}
\email{vittorio.giovannetti@sns.it}

\begin{abstract}
As the possibility to decouple temporal and spatial variations of the electromagnetic field, leading to a wavelength stretching, has been recognized to be of paramount importance for practical applications, we generalize the idea of stretchability from the framework of electromagnetic waves to massive particles.
A necessary and sufficient condition which allows one to identify energetically stable configuration of a 1D  quantum particle characterized by arbitrary large spatial regions where
the associated wave-function exhibit  a  flat, non-zero  profile is presented, together with examples on well-known and widely used potential profiles and an application to 2D models.
\end{abstract}

\maketitle
\section{Introduction}

In recent years much attention has grown around the possibility of developing
photonic metamaterial with {\it near-zero parameters} (for instance media with near-zero relative permittivity and/or relative permeability, which imply near-zero refractive index)~\cite{meta_review}.
This interest is due to the peculiar effects and  applications that such artificial structures allow for: in near-zero refractive index media electricity and magnetism decouple even at non-zero frequency, leading to a corresponding effective  decoupling of spatial and temporal field variations~\cite{pre70046608,science340}
which enables for wave profiles having both large frequencies and large (stretched) wavelengths.
The independence of wavelength and frequency has great importance from both the theoretical  and the technological perspective: many effects have been foreseen and some have already been verified experimentally. Among them we cite tunneling through distorted channels~\cite{prl97157403,prl100033903,jap105044905}, highly directive emitters~\cite{prl89213902}, radiation pattern tailoring~\cite{prb75155410,apl105243503}, boosted non-linear effects~\cite{prb85045129,prb89235401,phase_mismatch,dipole_emitter} and cloaking~\cite{cloaking_theo,cloaking_exp}.
In the uncorrelated field of condensed matter physics, artificial structures with engineered bands profile has been investigated since the first proposal for superlattices by Esaki and Tsu~\cite{esaki}: 
from then on many progresses have been made exploiting band engineering~\cite{revmodphys73783}, leading to the realization of 2-dimensional electron gases, quantum wires~\cite{beenakker}, and dots~\cite{repprogphys64701,rmp741283}. Furthermore, relaying on the formal analogies that, via the particle-wave
duality, links photons 
 and electrons~\cite{datta}, 
some pioneering works have started investigating what 
metamaterials can bring to the field of semiconductor physics.
 In this context many proposals have been done, from matter waves subwavelength focusing~\cite{prl110213902} and matter waves cloaking~\cite{prl100123002,prb88155432} to spintronics applications~~\cite{chesi}, just to make few examples. Also new devices have been proposed exploiting the electron-photon analogy, such as superconducting structures~\cite{science328582,natphys4929}, faster integrated circuits and optical devices~\cite{prb86161104,prb89085205} and connectors for misaligned channels~\cite{prb90035138}.

Inspired by these approaches in the present work we discuss about the possibility of 
producing energetically stable configurations for confined  massive particles,  characterized by 
 wavefunctions that exhibit extended flat non-zero spatial regions with almost zero associated momentum.
We dub these special states
{\it stretched quantum states} as they  possess some analogies with the stretched electromagnetic waves.
In our construction we assume the possibility of carefully  tailoring the confining potential that traps the particle. Focusing hence on the paradigmatic case where the dynamics is effectively constrained along a 1D line, we provide necessary and sufficient conditions that univocally 
identify the set of {\it stretching potentials}, i.e. the set of potentials which admit a stretched quantum state
among their eigenfunctions.

The paper goes as follows: in Sec.~\ref{sec:solutions} we set the problem and present a general 
construction to realize stable stretched configurations for a massive, non-relativistic 1D particle. 
In Sec.~\ref{sec:high} we discuss about possible generalization to higher spatial dimensions,
discussing in particular an application for 2D scattering models. 
In Sec.~\ref{sec:examples} we present some explicit examples of stretching potential discussing their
spectral properties. 
Finally in Sec.~\ref{sec:conclusions} we draw our conclusions.


\section{\label{sec:solutions}Stretched energy eigenstates and stretching potentials}

\begin{figure}[t]
\includegraphics[scale=0.5]{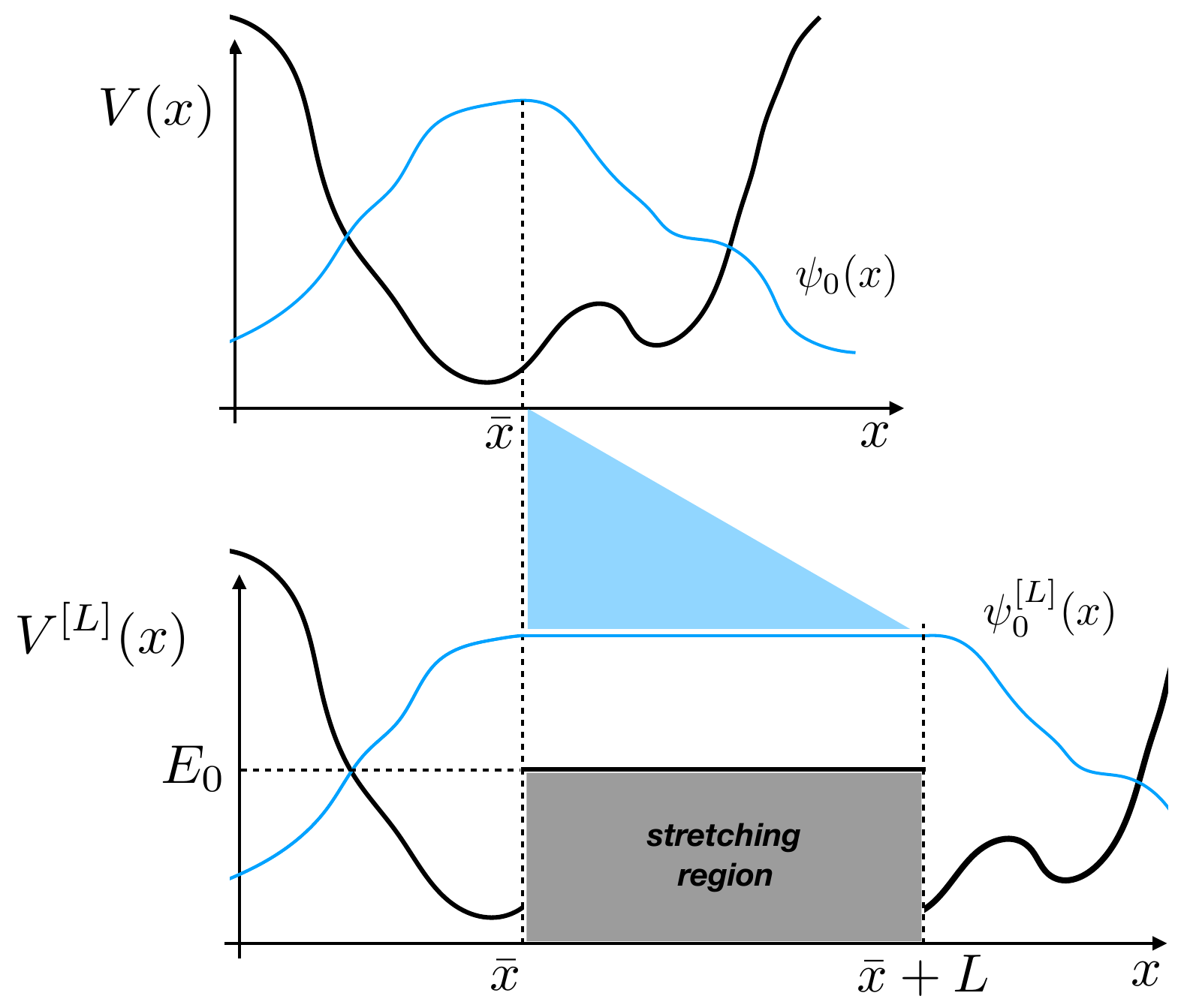}
\caption{Pictorial  sketch of the wave-function stretching procedure.
Top panel: the original potential trap $V(x)$ (black curve) and the associated wave-function $\psi_E(x)$ (blue curve)
relative to the energy level $E$; Bottom panel: 
modified potential profile $V_0^{[L]}(x)$ obtained by cutting $V(x)$ into halves in correspondence of the stationary
point $\bar{x}$ of $\psi_0(x)$, connecting the two parts with a constant profile of value equal to $E$; the blue curve
represents the value of new associated wave function $\psi_0^{[L]}(x)$ which presents a stretched region in the interval
$[\bar{x}, \bar{x}+L]$. 
}
\label{FIG1}
\end{figure}

It is a well known fact that the Schr\"odinger eigenvalue equation for 
a non-relavistic massive particle,
\ba\label{SCH} 
\nabla^2\psi_E(\vec{x}) &=&-\frac{2m}{\hbar^2}(E-V(\vec{x}) )\psi_E(\vec{x}) \;, 
\ea
 bares a close similarity with the Helmholtz equation for the electro-magnetic field, the latter being
 formally obtained from (\ref{SCH}) by replacing the wave-function $\psi_E(\vec{x})$ with 
 the electro-magnetic field $\vec{E}(\vec{x})$ and identifying $\frac{2m}{\hbar^2}(E-V(\vec{x}) )$
 with the term $\omega^2\mu\epsilon$, $\omega$ being the frequency of the signals, 
 $\epsilon$ and $\mu$ being instead  the permittivity  and permeability of the medium. 
In photonic metamaterials one between these last two quantities  is artificially set to zero, leading to an effective  decoupling of spatial and temporal variations of the field and to an infinite phase velocity. One immediately recognizes that the analogous condition for matter waves is to have $E-V(\vec{x})$ equal to zero.
More precisely adopting a reverse engineering point of view, we can use  Eq.~(\ref{SCH}) as a tool
for identifying the spatial properties of the potential $V(\vec{x})$
that allows one to promote a generic target wave function $\psi_{\text{tar}} (\vec{x})$ (i.e. the wave function that we aim to obtain) to an energetically stable configuration
 of the model, i.e. 
\ba\label{INVSCH} 
V(\vec{x}) - E = \frac{\hbar^2}{2m}  \frac{\nabla^2\psi_{\text{tar}}(\vec{x})}{\psi_{\text{tar}}(\vec{x})} \;,
\ea
with $E$ playing the role of a free parameter that we can fix at will. Accordingly, requiring $\psi_{\text{tar}}(\vec{x})$ to assume a constant, non-zero value on a spatial domain ${\cal D}$,  Eq.~(\ref{INVSCH})  
can be used as a tool to identify the corresponding stretching potential.  
In particular from it we can estrapolate that  a necessary condition that such special 
 $V(\vec{x})$ must fulfil   is the fact that it  has to assume  
constant value on ${\cal D}$, i.e.
\begin{eqnarray} 
\psi_{\text{tar}}(\vec{x}) = {const.} &\quad&  \forall \vec{x} \in {\cal D}\;,\nonumber  \\
&& \Longrightarrow \label{COND1} 
V(\vec{x}) =  {const.}  \quad \forall \vec{x} \in {\cal D}\;. 
\end{eqnarray} 
Reversing the  implication of Eq.~(\ref{COND1}) 
is clearly a much more subtle problem: indeed, due to the need of properly matching boundary conditions,
 there is  no guarantee that a given potential $V(\vec{x})$  that
is constant on certain domain ${\cal D}$ will be also a stretching potential.
 In what follows we shall focus on this specific task presenting a 
 solution to the problem based on a simple reshaping of the potential  
 which allows one to create energetically stable spatially-flat orbits from
 energetically stable non-flat ones. 
 In the presentation that follows  
we shall specifically address 
 the  paradigmatic case where the particle is confined along a
1D line for which an explicit analytical treatment is allowed, and for which the strategy we propose results to be necessary and sufficient for characterizing the set of stretching potentials.

\begin{figure}[t]
\includegraphics[scale=0.5]{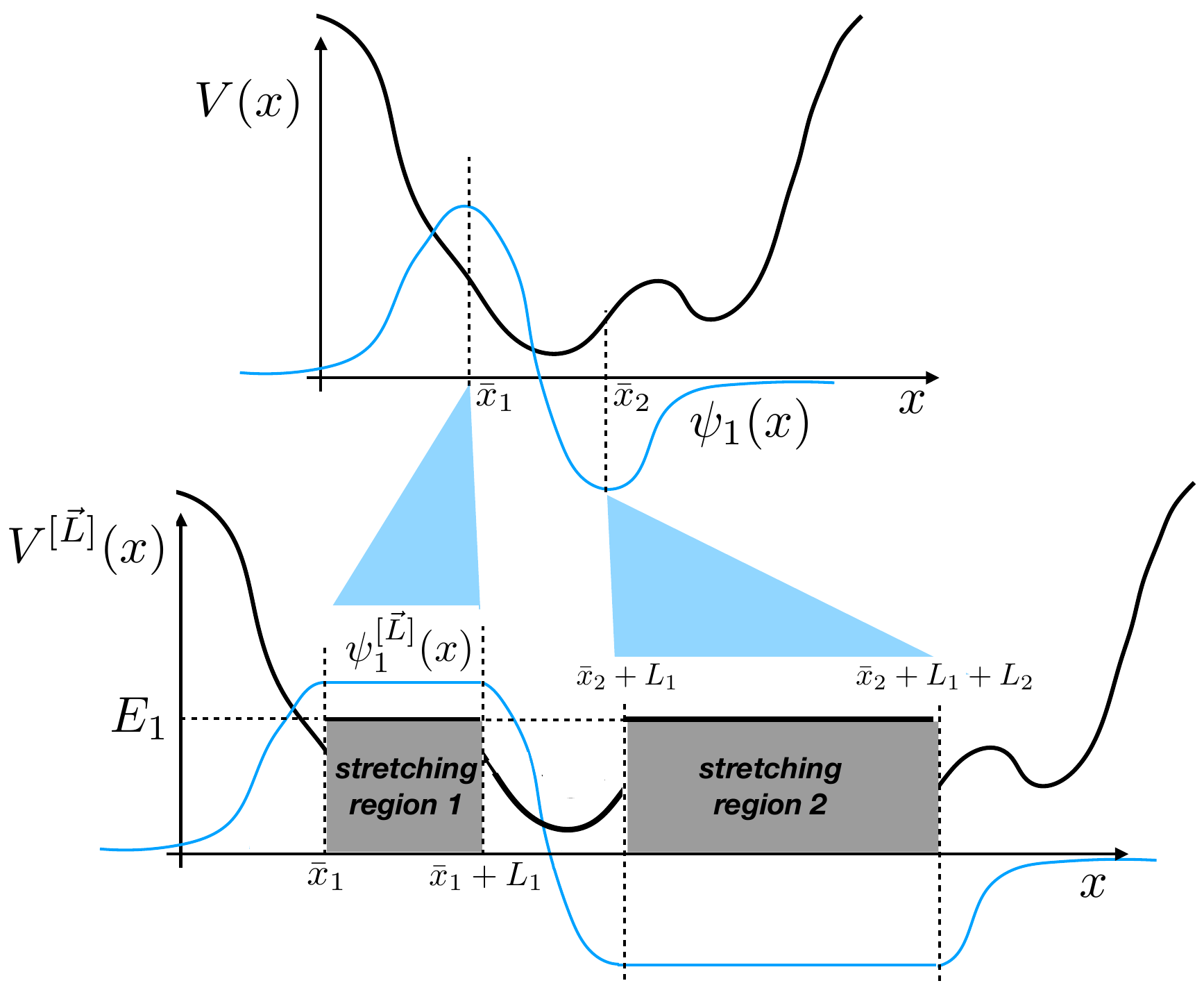}
\caption{Pictorial representation of a two-point stretching procedure for the first excited state of the model.
Top panel: original potential trap $V(x)$ (black curve) and the associated wave-function $\psi_1(x)$ (blue curve)
relative to the  first excited energy level $E_1$; Bottom panel: 
modified potential profile $V^{[\vec{L}]}(x)$ (black curve). Notice that this time the cutting points are two in correspondence 
of the stationary  positions 
 $\bar{x}_1$ and $\bar{x}_2$ of $\psi_1(x)$. 
}
\label{FIG2}
\end{figure}

\subsection{Constructing stretching potentials for a massive 1D particle} \label{sec1} 
Consider  a non-relativistic massive particle $A$ 
obeying the Schr\"{o}edinger equation 
\begin{eqnarray}
\label{eq:standard_schroedinger}
\partial^2_x \psi_E({x}) &=&-\frac{2m}{\hbar^2}(E-V({x}) )\psi_E({x}) \;,
\end{eqnarray}
with  $V(x)$ that we dub {\it seeding} potential.
From general results of  Sturm-Liouville equations, given a time-reversal invariant Hamiltonian with a non-degenerate spectrum, it is known that the 
 bound-state solutions of Eq.~(\ref{eq:standard_schroedinger}) form an orthonormal set of eigenfunctions with associated eigenvalues $\{\psi_n(x), E_n\}_{n=0,1,2, \dots}$ which we may assume to be labelled in (strict) energetically increasing  order by the index $n$.
We also know that, under our hypotheses, the $n$-th eigenfunction $\psi_n(x)$ can be chosen to be real, and that it must have exactly $n$ nodes,
 and thus at least $n+1$ spatially separated extremal points
  in the domain of definition of the problem. 
In particular let us  indicate with $\bar{x}$ the position of one of the extremal points of 
the ground state  $\psi_0(x)$ of Eq.~(\ref{eq:standard_schroedinger}), i.e. 
$\partial_x\psi_0(\bar{x})=0$. 
Consider hence the following modification of the confining potential  obtained by 
``cutting" into halves the seeding potential ${V}(x)$ at point $\bar{x}$, separating them by a spatial distance ${L}\geq 0$
and introducing an intermediate step-like potential of value equal to the original ground state
energy level $E_0$, i.e. 
\begin{eqnarray}
{V}^{[{L}]}(x):=\left\{\begin{array}{ccc}
{V}(x) & &\mbox{for $x\leq\bar{x}$,}\\
&&\\
E_0& &\mbox{for ${\bar{x}}< x<\bar{x}+{L}$,}\\
&&\\
{V}(x-{L})& & \mbox{for $x\geq\bar{x}+{L}$,}
\end{array}\right.
\label{NEWPOT} 
\end{eqnarray}
see Fig.~\ref{FIG1}. 
The crucial observation is that the associated modified 
Schr\"odinger equation
 \begin{eqnarray}
\label{eq:standard_schroedingerNEW}
\partial^2_x \psi_E({x}) &=&-\frac{2m}{\hbar^2}(E-{V}^{[{L}]}({x}) )\psi_E({x}) \;,
\end{eqnarray}
still admits $E=E_0$ as eigenvalue for all possible choices of the stretching parameter ${L}$. Indeed  
  in the region $x\leq\bar{x}$, 
  an explicit solution $\psi^{[{L}]}_{E=E_0}({x})$ of Eq.~(\ref{eq:standard_schroedingerNEW}) for 
 $E=E_0$ can be obtained by taking it equal to  $\psi_{0}({x})$ of 
Eq.~(\ref{eq:standard_schroedinger}). 
 Similarly in the region $x\geq\bar{x}+{L}$, we can take as $\psi^{[{L}]}_{E=E_0}({x})$
 the translated version   $\psi_0(x-{L})$  of $\psi_0(x)$.
 We are thus left with the central region of length ${L}$ where the potential is constant and equal to $E_0$: here 
 the modified Schr\"odinger equation 
 admits $E_0$ as possible solution once we take $\psi_E({x})$ to be constant, e.g. 
equal to {${\psi}_0(\bar{x})$} to match the necessary boundary conditions.
Recapping, starting from the seeding potential $V(x)$ which in principle may have no stretched eigenstate at $E_0$,   we have identified a one parameter family of {stretching-}potential profiles~
\begin{eqnarray} {\cal F}^{[L]}(V):= \{ 
V^{[{L}]}({x}); L\geq  0\}\;, \end{eqnarray} 
 whose $L$-element 
admits, up to an irrelevant normalization prefactor,  the  {\it stretched}  wave-function 
\begin{eqnarray} 
\label{NEWSOLU} 
\psi^{[{L}]}_0(x):=\left\{\begin{array}{ccc}
\psi_0(x) & &\mbox{for $x\leq\bar{x}$,}\\
&&\\
{{\psi}_0(\bar{x})}& &\mbox{for $\bar{x}< x< \bar{x}+{L}$,}\\
&&\\
\psi_0(x-{L})& & \mbox{for $x\geq\bar{x}+{L}$,}
\end{array}\right.
\end{eqnarray}
as ground state eigenvector associated with the {\it same} eigenvalue $E_0$ of the original (un-modified, ${L}=0$) potential 
$V(x)$ -- the condition implied by Eq.~(\ref{COND1}) being of course fulfilled by all the elements of the family. 
Essentially we can say that the above construction allows one to create an extended spatial region where 
 all the kinetic  energy component of the 
wave-function is smoothly converted into potential energy without modifying the total energy eigenvalue.

By construction the states  (\ref{NEWSOLU})  possess the same number (i.e. zero) of nodes as $\psi_0(x)$: accordingly 
 $\psi^{[{L}]}_0(x)$ must represent the ground state of the
new Hamiltonian model, making $E_0$ the ground state energy level of the modified scheme irrespectively from the
chosen value of the stretching parameter ${L}$, i.e. 
\begin{eqnarray} \label{STRECTH} 
E_0^{[{L}]} = E_0\;, \qquad \forall {L}\geq 0\;. 
\end{eqnarray} 
The above observations of course
 do not apply to the other 
energy levels of the  seeding potential. Namely i) for $L>0$, the other energy eigenvectors 
 of the Hamiltonian associated with $V^{[{L}]}(x)$ will not correspond to stretched versions of their 
 original counterparts; and  ii) their associated eigenvalues will not coincide with their original
 counterparts. Instead, 
 as the potential (\ref{NEWPOT}) 
is more shallow than the original one, one expects the energy gaps between the
excited eigenvalues $E_n^{[{L}]}$ of the new Hamiltonian to get reduced as ${L}$ increases, i.e. 
\begin{eqnarray}
E_{n+1}^{[{L}]} - E_n^{[{L}]} \leq E_{n+1} - E_n \;, \end{eqnarray}  
and to vanish in the asymptotic infinite stretching  limit ${L}\rightarrow \infty$,
resulting into an overall compression of the energy spectrum 
(explicit evidences of this behavior will be provided in the next sections).

In case the ground state wave-function $\psi_0(x)$ of the seeding potential $V(x)$, 
possesses more than a single, say $j>1$, stationary points, 
the same construction can be repeated to each one of them independently 
leading to the identification of a larger family of stretching
potentials 
\begin{eqnarray} {\cal F}^{[\vec{L}]}(V):= \{ 
V^{[{\vec{L}}]}({x}); \vec{L}:= (L_1, L_2, \cdots, L_{j})  \}\;, 
\label{NFAMI} \end{eqnarray} 
characterized  now by $j$ positive independent parameters $L_1, L_2, \cdots, L_j$, each inducing 
a different, independent modification on $V(x)$. 
Specifically in this new scenario $V^{[{\vec{L}}]}({x})$ is 
obtained by cutting the seeding potential  into $j+1$ pieces in correspondence to the stationary points $\bar{x}_1, {\bar{x}_2}, \cdots, \bar{x}_{j+1}$ of $\psi_{0}(x)$, separating the various terms
by intervals of  lengths specified by  $L_1, L_2, \cdots, L_{j+1}$  respectively, upon which $V^{[\vec{L}]}(x)
$  is set constant and  equal to  $E_0$. 
The corresponding  modified eigenfunction $\psi^{[\vec{L}]}_0(x)$ can then be constructed along the same line
reported in (\ref{NEWSOLU}), and by using the same zero-nodes counting arguments adopted previously one can again 
show that, for fixed values of $\vec{L}$,  it will be the ground state 
 configuration of the model -- its energy being still equal to $E_0$, all the other energy levels of the model being instead squeezed toward it. 

More generally by following the same construction detailed above,
stretched versions of the excited energy levels of the original Hamiltonian 
can also be obtained: for instance in the case of the $n$-th eigenlevel  $\psi_n(x)$,
we shall now use as cutting points for the seeding potential the stationary points of such wave-function,
and set equal to $E_n$  the constant value of the associated elements $V^{[{\vec{L}}]}({x})$ of the 
stretching potential  family~ ${\cal F}^{[\vec{L}]}(V)$  (see Fig.~\ref{FIG2}) -- notice that in this case the dimension $j$ of the
vector $\vec{L}$ is larger than or equal to $n+1$ (the latter being the minimum number of stationary points of 
 $\psi_n(x)$).
Thus, by exploiting once more the node-counting argument 
one can finally conclude that in this case, 
$E_n$ represents the $n$-th energy bound state of the new Hamiltonian, i.e.
\begin{eqnarray} 
E_n^{[\vec{L}]} = E_n\;, \qquad \forall {L}_1,L_2, \cdots \geq 0\;, 
\end{eqnarray} 
and that  for all $n'\geq n+1$, the energies gaps  connecting the energy levels 
 $E_{n'}^{[\vec{L}]}$  will get compressed as the length of vector  of $\vec{L}$  increases. 
The behaviour of the lower part of the spectrum on the contrary will typically be reacher than what observed in the case of the 
ground-state stretching,   and will strongly depend on the specific properties of the seeding potential $V(x)$: as a general rule 
one can anticipate however that in the infinite stretching limit, it will tend to produce
multiplex  of almost degenerate levels.
\\
\\
\subsection{A necessary and sufficient condition for 1D stretching} 
The simple construction we have presented in the previous section  is  rather general and, at {least} for the
case of 1D geometries, 
 allows for an exhaustive 
characterization of stretching potentials. Indeed given a stretching potential $W(x)$ admitting a stretched state $\psi_E(x)$ as  eigenvector associated with one of its  eigenvalues $E$, then it turns out that it must be  constructed from a seeding potential $V(x)$  having an explicitly non stretched eigenstate at that same energy level, via the mechanism detailed above. In other words, $W(x)$ must be an element of a  stretching family~(\ref{NFAMI}) characterized by a number of parameters $j$ 
that is larger than or equal to  $n+1$ with $n$ being the spectral position of the energy level $E$ of $W(x)$. The proof of this statement
can be constructed by reversing the various steps we have followed before. For instance, assume  that 
$E$ coincides with the ground state level of $W(x)$ and  that it admits a single stretching, fully connected, interval ${\cal I}:= [ \bar{x}, \bar{x}+\ell]$
of length $\ell>0$, i.e. 
\begin{eqnarray}
\psi_E(x) = const. &\quad& \forall x\in {\cal I} \;,  \nonumber \\
&& \Longrightarrow \label{COND111} 
W({x}) =  {const.}  \quad \forall {x} \in {\cal I}\;,
\end{eqnarray} 
 the condition~(\ref{COND111}) being a rewriting of  Eq.~(\ref{COND1}). 
 Given then $\ell' \in ]0,\ell]$, construct  a new potential profile ${W}^{[\ell']}(x)$ obtained from $W(x)$ by 
 maintaining the same spatial dependence for all $x\leq \bar{x}+ \ell-\ell'$ and defined as the shifted version of $W(x)$, 
  $W(x+\ell')$ for all $x >  \bar{x}+ \ell-\ell'$, i.e. 
   \begin{eqnarray}
{W}^{[{\ell'}]}(x):=\left\{\begin{array}{ccc}
{W}(x) & &\mbox{for $x\leq \bar{x}+ \ell-\ell'$,}\\
&&\\
{W}(x+{\ell}')& & \mbox{for $x >  \bar{x}+ \ell-\ell'$.}
\end{array}\right.
\label{NEWPOT111} 
\end{eqnarray}
Now by construction an  eigenvector  of ${W}^{[{\ell'}]}(x)$ with eigenvalue $E$ is  provided by the function 
\begin{eqnarray} 
\label{NEWSOLUinv111} 
\psi_E^{[{\ell'}]}(x):=\left\{\begin{array}{ccc}
\psi_E(x) & &\mbox{for $x\leq \bar{x}+ \ell-\ell'$,}\\
&&\\
\psi_E(x+{\ell'})& & \mbox{for $x>\bar{x}+\ell-\ell'$,}
\end{array}\right.
\end{eqnarray}
the function fulfilling the proper boundary conditions thanks to the fact  that $\psi_E(x+{\ell'})$ and 
${W}^{[{\ell'}]}(x+\ell')$ 
are constant
for all
$x\in ] \bar{x}+\ell-\ell', \bar{x}+\ell]\subseteq {\cal I}$. 
We observe that while for all $\ell' <\ell$,  $\psi_E^{[{\ell'}]}(x)$ is a stretched state (being constant upon
a non-zero interval), this is no longer the case for $\ell' = \ell$ where ${W}^{[\ell]}(x)$ becomes  a seeding
potential with an eigenfunction $\psi_E^{[{\ell}]}(x)$ that, by construction, admits a stationary point in $\bar{x}$ and no stretching
elsewhere. 
The proof of the property we have stated above finally follows by noticing that
taking $V(x)= {W}^{[\ell]}(x)$, we can write $W(x)={V}^{[\ell]}(x)$ hence showing that 
$W(x)$ belongs to the family ${\cal F}^{[L]}( {W}^{[\ell]}(x))$.
\\

\section{Stretching the wave-function in higher spatial dimensions} \label{sec:high} 

A rather obvious generalization of the results presented in Sec.~\ref{sec:solutions} can be obtained in the 2D or 3D settings,
for all those models whose seeding potentials exhibit an explicit separation between the various cartesian coordinates, 
such~as
\begin{eqnarray}\label{DDF} 
V(\vec{x}) =\sum_{j}  V_j(x_j) \;,
\end{eqnarray} 
with $x_j$ being the $j$-th coordinate of $\vec{x}$, the $V_j(\cdots)$ being arbitrary functions. 
Indeed under this assumption the Schr\"{o}edinger equation (\ref{SCH}) admits solutions of the form
\begin{eqnarray}
\psi_E(\vec{x}) = \Pi_j \psi^{(j)}_{E_j} (x_j) \;,\qquad E= \sum_j E_j\;,\end{eqnarray} 
where for all $j$,  $\psi^{(j)}_{E_j}(x)$ and $E_j$ satisfy  the identity
\begin{eqnarray}
\label{eq:standard_schroedingerJJJ}
\partial^2_x \psi^{(j)}_{E_j}({x}) &=&-\frac{2m}{\hbar^2}(E_j-V_j(x) )\psi^{(j)}_{E_j}({x}) \;.
\end{eqnarray}
Accordingly, by applying the procedure detailed in the previous section for each one of the 
wave-functions $\psi^{(j)}_{E_j}(x)$ independently,  we can produce stretched versions of $\psi_E(\vec{x})$.
An important point to be highlighted is that, in order to apply the procedure we illustrated in 1D, one must also require the boundary conditions on the wave function to be factorized.

\begin{figure}[!t]
\centering
\includegraphics[scale=0.3]{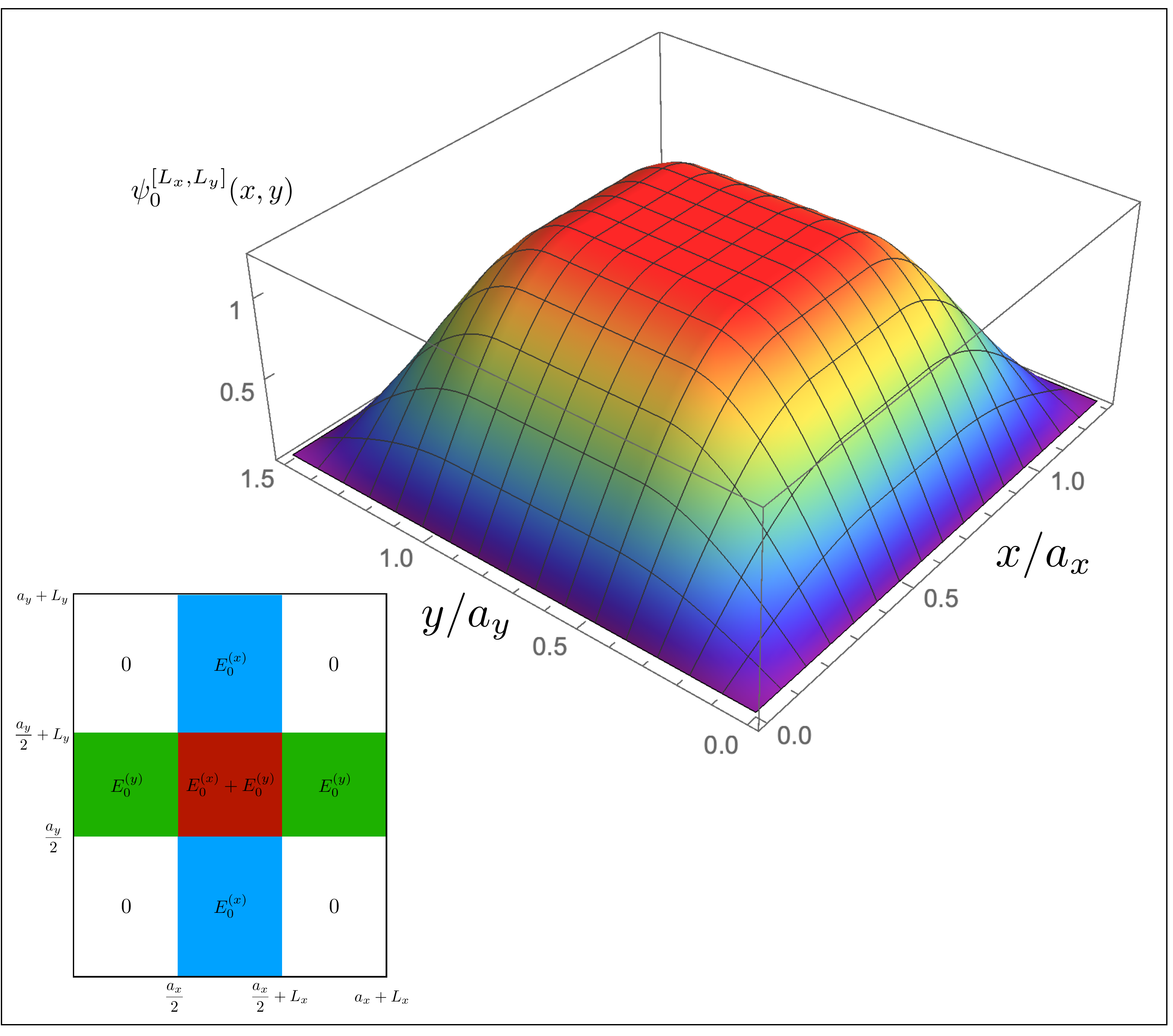}
\caption{Inset: {density plot of the 2D-stretching potential ~(\ref{2DSTREX}). Figure: plot of the 2D stretched ground state $\psi_0^{[L_x,L_y]}(x,y)$ for $L_x=0.3a_x$ and $L_y=0.5a_y$, see Eq.~(\ref{GROUND2D})}. One can observe the flat region at the center, in correspondence of the red region of the inset, where the wave function has a constant value. One can also observe four regions where only one component of the gradient nullifies, corresponding to the blue and green regions of the inset.}
\label{fig:2d_example}
\end{figure}

Consider for instance the case of an infinite 2D square well, that is a potential of the form:
\ba
V(x,y)=V_x(x)+V_y(y)
\ea
where for $j=x,y$ we set
\ba
V_j(j)=\left\{\begin{array}{cc}
0&\mbox{for $0\leq x_j\leq a_j$},\\
+\infty&\mbox{otherwise}, \label{2DSTREX} 
\end{array}\right.
\ea
with  $a_j$ the width of the well along the $j$ direction.
This potential is manifestly separable, and thanks to the fact that the potential is infinite at the boundaries, also the boundary conditions are separable. We want to consider the stretching of the ground state, which can be written as:
\ba
\psi_0(x,y)=\psi_0^{(x)}(x)\psi_0^{(y)}(y),
\ea
where
\ba
\psi_0^{(j)}(j)=\sqrt{\frac{2}{a_j}}\sin(\frac{\pi}{a_j}j),
\ea
whose associated eigenenergies are $E_0^{(j)}=\frac{\hbar^2\pi^2}{2ma_j^2}$, so that $E_0=E_0^{(x)}+E_0^{(y)}$.
From these expressions one immediately sees that the wave function $\psi_0(x,y)$ has zero gradient at the point $(\frac{a_x}{2},\frac{a_y}{2})$: we hence insert a potential box of height $E_0^{(x)}$ and width $L_x$ at the point $x=a_x/2$, and similarly a box of height $E_0^{(y)}$ and width $L_y$ at point $y=a_y/2$, obtainig the potential shown in the inset of Fig.~\ref{fig:2d_example}. This new potential ground state eigenfunction reads:
\ba
\psi_0^{[L_x,L_y]}=\psi_0^{(x)[L_x]}(x)\psi_0^{(y)[L_y]}(y),
\ea
where now $\psi_0^{(j)[L_j]}(j)$ is worth:
\ba
&&\frac{1}{\sqrt{N_j}}\left\{\begin{array}{cc}
\sin(\frac{\pi}{a_j}j)&\mbox{for $0\leq j\leq\frac{a_j}{2}$},\\
1&\mbox{for $\frac{a_j}{2}\leq j\leq\frac{a_j}{2}+L_j$},\\
\sin(\frac{\pi}{a_j}(j-L_j))&\mbox{for $\frac{a_j}{2}+L_j\leq j\leq a_j+L_j$}.
\end{array}\right. \label{GROUND2D} 
\ea
This leads to the ground state wave function plotted in Fig.~\ref{fig:2d_example}, where one has a central region where $\psi_0^{[L_x,L_y]}(x,y)$ is constant, and four regions where one of the components of the gradient is null. These four regions correspond to the blue and green regions of the inset of Fig.~\ref{fig:2d_example}, where only one the kinetic components is absorbed into potential energy.

Extending this construction beyond the cases included in Eq.~(\ref{DDF}) is much more demanding
due to the intrinsic interplay between the various coordinate components introduced by the seeding potential  which prevents us from
operating on one of them individually without affecting the others, leading in general to ambiguities affecting the boundary conditions which make not clear how to perform reverse engineering. 
  It turns however that  
 if we do restrict ourselves to 
small departure from the condition~(\ref{DDF}), approximate solutions can be found. 
To present this result we shall focus on a 2D geometry for which 
 practical applications  can be envisioned in the design of 
semiconductor electronic wave-guides. For this purpose let us consider 
a particle
$A$ that moves on the $(x,y)$ plane  under the action 
of a seeding potential which is translationally invariant along the longitudinal $y$ direction, i.e. $V(x,y) =V(x)$, hence writing (\ref{SCH}) as
\begin{equation}
\label{eq:standard_schroedinger2D}
(\partial^2_x + \partial^2_y )\psi_E({x,y}) =-\frac{2m}{\hbar^2}(E-V({x}) )\psi_E({x,y}) \;.
\end{equation}
The model clearly admits eigen-solutions of the form 
\begin{equation} \label{ans1new} \psi_E({x,y})= e^{i ky} {\psi}_{n}(x) \;, \quad 
E= E_n +  \hbar^2 k^2/2m\;, \end{equation} 
for $k$ real,  and ${\psi}_{n}(x)$  being the $n$-th eigenstate 
of the 1D problem defined by  $V(x)$ and associated with the eigenvalue $E_n$
(in what follows we shall assume this part of the spectrum to be discrete).
Now considering $n=0$,  i.e. identifying ${\psi}_{0}(x)$ and $E_0$ with the ground energy
level associated with the 1D potential $V(x)$, let us consider the following modification of
(\ref{eq:standard_schroedinger2D})  \begin{equation}
\label{defimpo} 
(\partial^2_x + \partial^2_y )\psi_E({x,y}) =-\frac{2m}{\hbar^2}(E-V^{[{L}(y)]}(x) )\psi_E({x,y}) \;,
\end{equation}
where {$V^{[{L}(y)]}(x)$} is obtained as in Eq.~(\ref{NEWPOT}) when identifying the 1D seeding
potential of that equation with the $V(x)$ appearing in 
Eq.~(\ref{eq:standard_schroedinger2D}), and where
${L}(y)$ is a (positive) smooth function of the longitudinal coordinate~$y$.
In what follows we shall consider the case where ${L}(y)$ 
 varies only on a limited
spatial interval $y\in {\cal I} :=[y_{in}, y_{fin}]$, assuming the constant value $L_{in}:=L(y_{in})$ for $y\leq y_{in}$ and $L_{fin}:=L(y_{fin})$ for $y\geq y_{fin}$. 
For the special choice in which the two asymptotic values coincide (i.e. $L_{in}= L_{fin}=L$),  
{and} ${L}(y)$ is constant and the model retains its invariance under  longitudinal translations ($V(x,y)=V^{[{L}]}(x)$): accordingly 
the solutions of
(\ref{eq:standard_schroedinger2D}) can be still obtained by separation of the coordinates through the
ansatz 
\begin{equation} \label{ans1new} \psi_E({x,y})= e^{i ky} {\psi}^{[L]}_{n}(x) \;, \quad 
E= E^{[L]}_n +  \hbar^2 k^2/2m\;, \end{equation} 
where now ${\psi}^{[L]}_{n}(x)$ and $E_n^{[L]}$ are  eigensolutions 
of the 1D problem characterized by the 1D potential $V^{[{L}]}(x)$.
For $n=0$, Eq.~(\ref{defimpo}) exhibits in particular a modification of 
$e^{i ky} {\psi}_{0}(x)$ that, for all assigned $y$, is uniformly stretched along the transverse direction $x$, i.e.  
the wave-function 
\begin{equation} 
\psi_E(x,y) = e^{i ky} \psi^{[{L}]}_0(x) \;, \quad 
E= E_0 +  \hbar^2 k^2/2m \;, \label{EASYwq} 
\end{equation} 
that simply provides  an instance of the construction we have anticipated at the beginning of the present section.
The situation changes however when we allow for arbitrary choices of $L_{in}$ and $L_{fin}$.
In this case   clearly the above construction 
fails since the potential (\ref{defimpo}) will acquire an explicit dependence on $y$. 
Still  a useful way to approaching the problem 
is to consider a modifed version of the 
ansatz (\ref{ans1new}) 
\begin{eqnarray} 
\psi_E(x,y) = e^{i ky} {\tilde{\psi}(x,y)} \;,
\end{eqnarray} 
where now, apart from the phase term $e^{i ky}$, we allow for a residual $y$-dependence in 
{$\tilde{\psi}(x,y)$}.
By replacing this into (\ref{defimpo}) 
  we get  
\begin{equation}
\label{eq:standard_schroedinger2Dnew}
\partial^2_x \tilde{\psi} (x,y)=-\frac{2m}{\hbar^2}\big[ E-V ^{[{L}(y)]}(x) \big] \tilde{\psi}(x,y)+  \Delta(x,y) \;, 
\end{equation}
with 
\begin{eqnarray} \label{dde} 
\Delta(x,y) : = - \partial^2_y \tilde{\psi}(x,y)- 2i k \partial_y \tilde{\psi}({x,y})\;.
\end{eqnarray} 
We reconize that apart from the $\Delta(x,y)$ contribution, Eq.~(\ref{eq:standard_schroedinger2Dnew}) 
reduces to   Eq.~(\ref{eq:standard_schroedingerNEW}) upon substituting 
$\tilde{\psi}({x,y})$ with $\psi_E({x})$.
Accordingly
one may try to use as an approximate  solution of (\ref{eq:standard_schroedinger2Dnew}) 
 the function
 $\psi^{[{L}(y)]}_0(x)$ obtained by evaluating  the 1D wave-function of 
 Eq.~(\ref{NEWSOLU}) for the stretching parameter ${L}(y)$, and taking for 
 $E$ the  corresponding ground state energy $E_0$, i.e. 
 \begin{equation} 
\psi_E(x,y) \simeq  e^{i ky} \psi^{[{L}(y)]}_0(x) \;, \quad 
E \simeq E_0 +  \hbar^2 k^2/2m \;. \label{EASY} 
\end{equation} 
This solution  turns out to be appropriate as long as
 we can neglect the $\Delta(x,y)$ contribution into Eq.~(\ref{eq:standard_schroedinger2Dnew}), a condition that, intuitively, 
 can be ensured 
if ${L}(y)$  
is a sufficiently slowly varying function with respect to $y$ on the whole interval ${\cal I}$ -- see Appendix ~\ref{APPA} for more details on this. 
Due to the presence of 
a non-constant stretching parameter $L(y)$, 
 Eq.~(\ref{EASY})  represents  a  generalization of the stretched state~(\ref{EASYwq}) 
 that applies for a model that, as anticipated, does not allow for trivial separation of variables.

\section{\label{sec:examples}Examples} 
Here we present few examples of the stretching mechanism for 1D models.
 \subsection{Infinite potential well}
As a study case we now choose as  
 seeding potential $V(x)$ an infinite well of length $a$ (i.e. 
$ {V}({x})=0$  for $|x|\leq\frac{a}{2},$ and ${V}({x})=\infty$ otherwise),  
which admits as energy eigenvectors the wavefunctions  supported in $|x|\leq\frac{a}{2}$
and defined by 
\begin{eqnarray}
\label{eq:inf_well_stand_sol}
\psi_n(x)=\sqrt{\frac{2}{a}}\left\{\begin{array}{cc}
\cos({\pi (n+1)x}/{a}) &\mbox{for $n\geq 0$ even,}\\
&\\
\sin({\pi (n+1)x}/{a})&\mbox{for $n\geq 1$ odd,}
\end{array}\right.
\end{eqnarray}
 with  associated eigen-energies equal to 
\begin{eqnarray}
\label{eq:infinite_well_eigenvalues}
E_n=\frac{\pi^2 (n+1)^2\hbar^2 }{2 m a^2}\;.
\end{eqnarray}

\subsubsection{\label{sec:gs_stretching}Ground state stretching}
\begin{figure}[!t]
\includegraphics[scale=0.75]{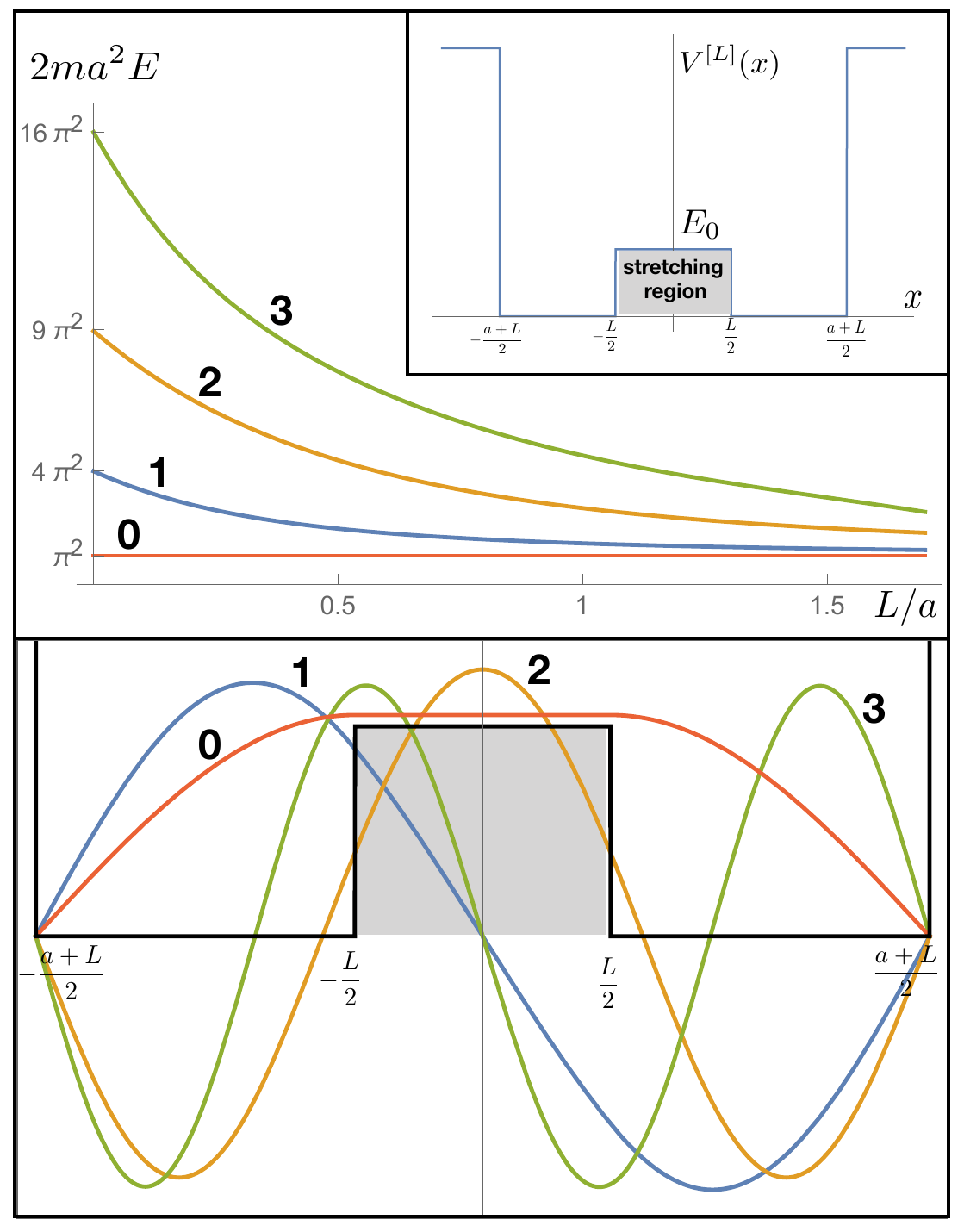}
\caption{Inset: A sketch of the potential profile under examination, an infinite well with a central barrier that gives rise to a stretched ground state. The height of the central barrier coincides with the energy ground state $E_0$ of the seeding
potential.
Upper panel: Energies of the first four levels ($n=0,1,2,3$)  of the potential $V^{[L]}(x)$ as a function of $L$. As $L=0$ one recovers the energy levels of the infinite well, while as $L$ increases these energies go asymptotically to the value $E_0$. In fact, in the limit of large $L$ one recovers the free particle limit, and thus the eigenfunctions become plane waves.
Lower panel: Plot of the wavefunctions of the first four levels (colors and numbers corresponding to the upper panel) for the specific ratio $L/a=0.4$. One can see immediately that the ground state wavefunction is flat in the region corresponding to the wall, while the other states are still oscillating eigenstates.}
\label{fig:inf_well_central_gs_levels}
\end{figure}

Observing that the ground state wavefunction $\psi_0(x)$ of the model has an extremal point at the origin of the coordinate axis, 
we can stretch 
it by considering 
the  potential profile shown in the inset of Fig.~\ref{fig:inf_well_central_gs_levels}, 
which at variance with the case presented in the previous paragraphs,  has been properly
shifted in order to ensure symmetry preservation  around $x=0$ for all $L$.
The explicit solution of the associated Sturm-Liouville equation~(\ref{eq:standard_schroedingerNEW}) can then be easily obtained by direct calculation. As anticipated in the 
previous section,
the first allowed solution is achieved for $E=E_0=\frac{\pi^2\hbar^2 }{2 m a^2}
$, the associated
wave-function being  provided by (\ref{NEWSOLU}).
The excited energy eigenfunctions for $E> E_0$ can  be obtained
by a standard approach. Setting 
$\bar{k}=\sqrt{\frac{2m(E-E_0)}{\hbar^2}}$ and ${k}=\sqrt{\frac{2mE}{\hbar^2}}$
 the resulting discrete spectrum emerges as the solution of the the following 
 quantization conditions
\begin{eqnarray}
\label{eq:inf_well_centrl_gs_quant_even}
k\cot\left(\frac{ak}{2}\right)=\bar{k}\tan\left(\frac{\bar{k}L}{2}\right)\;,
\end{eqnarray}
for states of even parity, and
\begin{eqnarray}
\label{eq:inf_well_centrl_gs_quant_odd}
k\cot\left(\frac{ak}{2}\right)=-\bar{k}\cot\left(\frac{\bar{k}L}{2}\right)\;,
\end{eqnarray}
for eigenstates of  odd parity.

In the upper panel of Fig.~\ref{fig:inf_well_central_gs_levels} we report the 
first four energy levels  as a function of the length $L$ of the central barrier, obtained by numerically solving the above expressions. As can be easily seen from the plot, 
while the ground state energy 
 is not affected by variations of $L$, 
 the excited levels have an explicit functional dependence on such parameter. 
In particular as $L$ is zero one recovers the infinite well eigenenergies, while as $L/a$ becomes large the excited
energy level get compressed toward
the ground state level $E_0$. 
The lower panel of Fig.~\ref{fig:inf_well_central_gs_levels} reports instead the eigenfunctions of the first
excited levels of the stretched potential for fixed choice of $L$ revealing the flat 
behaviour of $\psi_0^{[L]}(x)$.

\subsubsection{Excited state stretching by central potential}\label{sec:exc} 

Consider next the case where we modify the seeding potential to stretch one of its excited energy levels.  
For instance, exploiting the fact that 
a generic even eigenfunction of Eq.~(\ref{eq:inf_well_stand_sol}) still  admits a stationary point 
at $x=0$, we can stretch it 
by  using  the same symmetric profile given in the  inset of Fig.~\ref{fig:inf_well_central_gs_levels} by 
simply setting the value of the potential in the flat central region equal to the corresponding energy level $E_n$.
\begin{figure}[!t]
\includegraphics[scale=0.55]{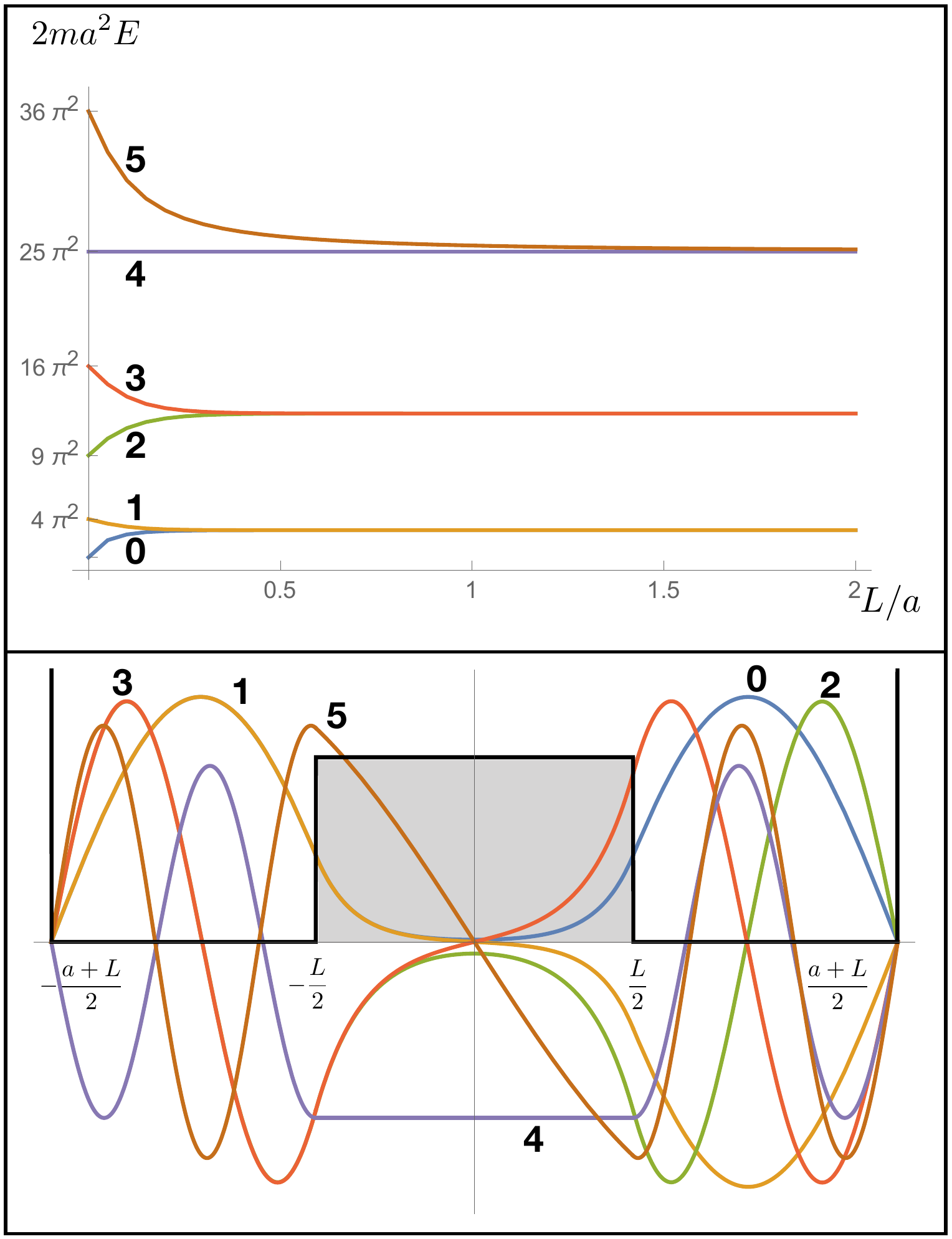}
\caption{Upper panel: Plot of the first six energy levels when the height of the central barrier is equal to the energy of the fifth level of the infinite well (i.e. $n=4$). As expected the energy of the fifth level keeps the same value as in the infinite well, independently of $L$. On the other hand the lower levels, as $L$ increases, acquire a doublets structure.
Lower panel: Plot of the wavefunctions of the first six levels for the specific ratio $L/a=0.6$. It is immediately seen that the doublets in the upper panel reflect here in couple of wavefunctions with opposite symmetry, i.e. even and odd states with the same energy value.}
\label{fig:inf_well_central_5s_levels}
\end{figure}
With such choice of course, irrespectively from the choice of the stretching parameter $L$,  $E=E_n$ turns out to be a proper eigenvalue of the modified Hamiltonian (indeed it is the $n$-th element of the spectrum). 
The eigensolutions for  $E\geq E_n$ can be solved as in the previous case and present
an analogous functional dependence upon $L$ (i.e. compression toward $E_n$ in the limit 
$L\gg a$). 
The system however now admits also energy levels for $E<E_n$ which,
setting $\gamma=\sqrt{\frac{2m(E_n-E)}{\hbar^2}}$, can be determined by 
solving the following quantization equations 
\begin{eqnarray}
\label{eq:inf_well_central_quant_cond_even}
-k\cot\left(\frac{ak}{2}\right)=\gamma\tanh\left(\frac{\gamma L}{2}\right)
\end{eqnarray}
 for even states,
 and
\begin{eqnarray}
\label{eq:inf_well_central_quant_cond_odd}
-k\cot\left(\frac{ak}{2}\right)=\gamma\coth\left(\frac{\gamma L}{2}\right)
\end{eqnarray}
for odd states.
 
 The above quantization conditions
 become exactly the same in the limit of $L\gg a$, and thus we expect the states with energy lower than $E_n$ to form two-fold nearly degenerate states with opposite parity.
The system can be seen as the union of two distinct potential wells separated by a finite barrier whose length acts as a knob that tunes the interaction between the two wells via tunnel effect. When the length $L$ is small the two wells interact strongly, while as $L$ increases the interaction becomes more and more feeble which gives doublets of nearly degenerate states. 
These predictions are confirmed by the plot shown in Fig.~\ref{fig:inf_well_central_5s_levels}. Now, a similar argument can be applied to the stretched state $n$=4 and the next higher state $n$=5: notice that the energy of the state $n$=5 in the limit of large $L$ approaches that of the stretched state and, correspondingly, its wave-function becomes practically linear in the stretched interval where the potential is constant and solutions with vanishing kinetic energy are of the form $\psi(x)=A$ (even symmetry) or $\psi(x)=B\,x$ (odd symmetry).

\subsubsection{\label{sec:multiparameter}Multi-parameter stretching of the first excited state}

Here we consider the multi-parameter stretching of the first excited state $\psi_1(x)=
\sqrt{\frac{2}{a}}\sin(2\pi x/a)$  of 
infinite potential well which admits $x=\pm a/4$ as stationary points. 
We hence use the following
stretched version of the seeding potential, i.e.   
\begin{eqnarray}
{V}^{[L_1,{L_2}]}(x)=\left\{\begin{array}{cc}
\infty&\mbox{for $x<-\frac{a}{2}-L_1$}\\
&\\
0&\mbox{for $-\frac{a}{2}-L_1\leq x\leq-\frac{a}{4}-L_1$}\\
&\\
E_1&\mbox{for $-\frac{a}{4}-L_1\leq x\leq-\frac{a}{4}$}\\
&\\
0&\mbox{for $-\frac{a}{4}\leq x\leq\frac{a}{4}$}\\
&\\
E_1&\mbox{for $\frac{a}{4}\leq x\leq\frac{a}{4}+{L_2}$}\\
&\\
0&\mbox{for $\frac{a}{4}+{L_2}\leq x\leq\frac{a}{2}+{L_2}$}\\
&\\
\infty&\mbox{for $x>\frac{a}{2}+{L_2},$}
\end{array}\right.
\end{eqnarray}
with $E_1= \frac{4 \pi^2 \hbar^2 }{2 m a^2}$ and $L_1,{L_2}$ being the two stretching parameters of the
problem. 
The energy levels can once more be easily computed. A part from the solution $E=E_1$ in
this case we find 
the following quantization condition
\begin{widetext}
\begin{eqnarray}
\nonumber
\frac{1}{4k^2\gamma^2\cos(k(\frac{a}{2}+L_1))}\Bigg\{-2\sinh(\gamma(L_1+{L_2}))k\gamma\Big((k^2+\gamma^2)\cos(\frac{ak}{2})
 +(k^2-\gamma^2)\cos(ak)\Big) &&  \\ 
  +\sinh(\gamma L_1)\sinh(\gamma {L_2})\sin(ak)(k^2-\gamma^2)^2 +2\sinh(\gamma {L_2})\sinh(\gamma L_1)\sin(\frac{ak}{2})(k^4-\gamma^4) &&\nonumber \\
  -4k^2\gamma^2\cosh(\gamma {L_2})\cosh(\gamma L_1)\sin(ak)  \Bigg\}=0&&\;, 
\end{eqnarray}
for $E<E_1$, where now $\gamma=\sqrt{\frac{2m(E_1-E)}{\hbar^2}}$ and $\bar{k}=\sqrt{\frac{2m(E-E_1)}{\hbar^2}}$, 
and 
\begin{eqnarray}
&&\frac{1}{8k^2\bar{k}^2\cos(k(\frac{a}{2}+L_1))}\Bigg\{-(k^2-\bar{k}^2)^2\cos(\bar{k}({L_2}-L_1))\sin(ak)+(k^4+6k^2\bar{k}^2+\bar{k}^4)\cos(\bar{k}(L_1+{L_2}))\sin(ak)\\
\nonumber
&&-4(k^4-\bar{k}^4)\sin(\frac{ak}{2})\sin(\bar{k}L_1)\sin(\bar{k}{L_2})+4k\bar{k}\sin(\bar{k}(L_1+{L_2}))\left[(k^2-\bar{k}^2)\cos(\frac{ak}{2})+(k^2+\bar{k}^2)\cos(ak)\right]\Bigg\}=0\;,
\end{eqnarray}
\end{widetext}
for $E>E_1$.
\begin{figure}[!t]
\includegraphics[scale=0.62]{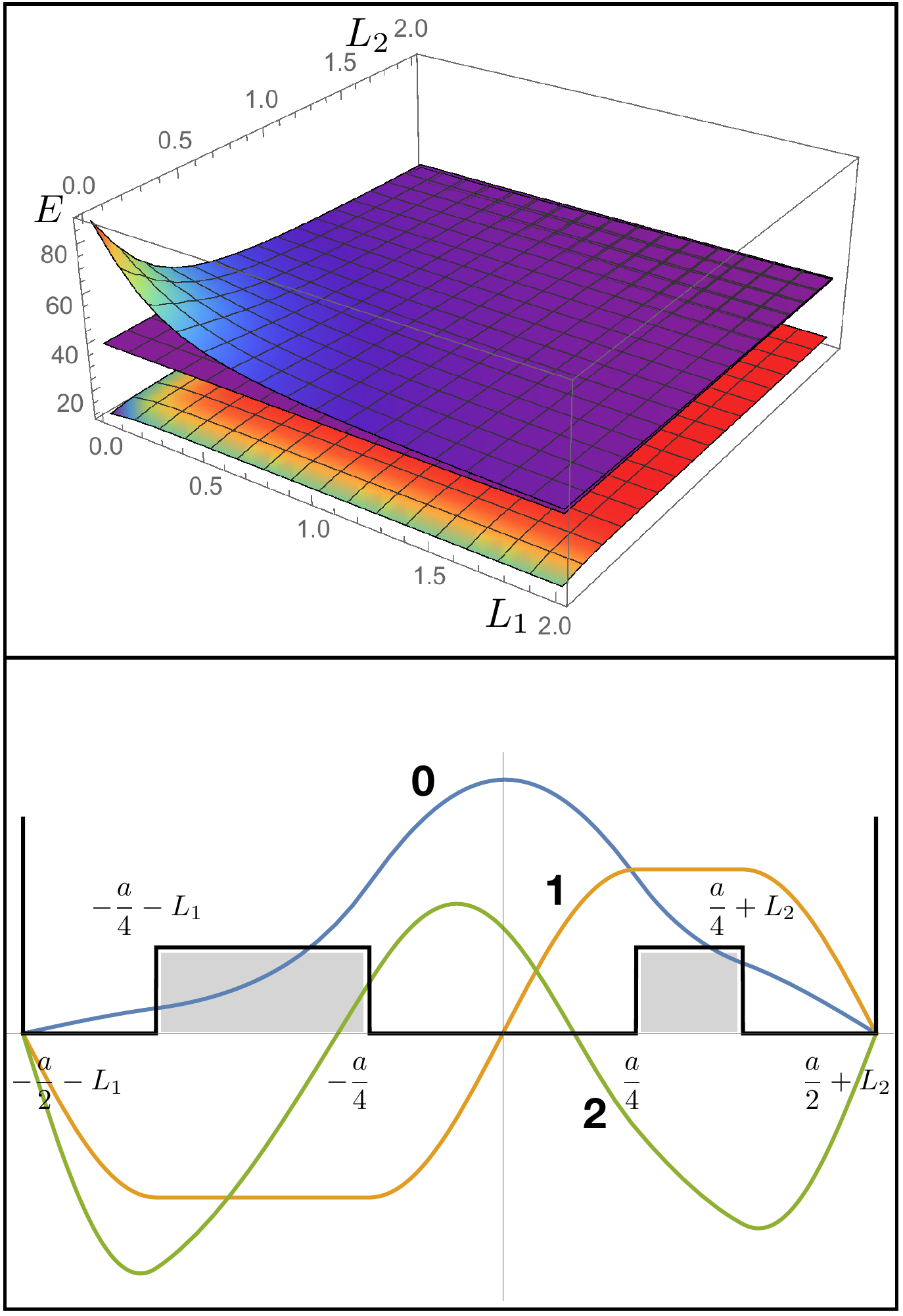}
\caption{Upper Panel, plot of the first energy levels for the infinite well with two barriers as a function of the barriers lenghts $L_1,{L_2}$. Lower Panel: plot of the wave-functions of the first three energy level of the model for $L_1/a=0.4$, $L_2/a=0.2$.}
\label{fig:en_levels_lateral}
\end{figure}

Plots of the first three  solutions as a function of 
are reported in the upper panel of Fig.~\ref{fig:en_levels_lateral}
 as a function of $L_1,{L_2}$:, one can see that the energy of the ground state increases up to an asymptotic value, while the energy of the second level stays constant, since it is the state we are stretching. As for the states with energy above $E_2$, one can see that their energy approach the asymptotic value $E_1$ as $L_1,{L_2}$ increase: this is to be expected, as we are going towards the free particle limit.

\subsection{Example 2: Harmonic oscillator}
\begin{figure}[!t]
\includegraphics[scale=0.55]{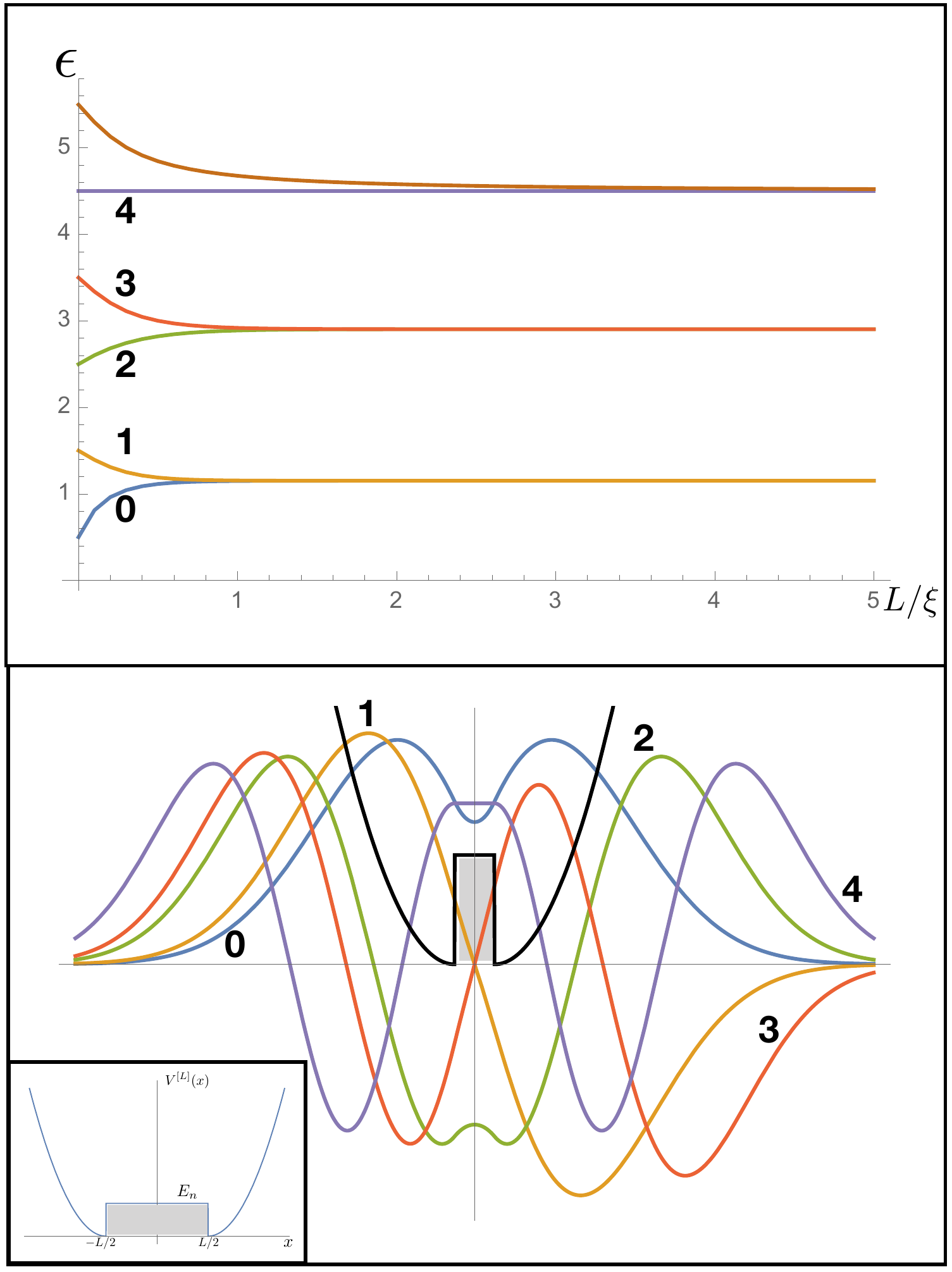}
\caption{Upper panel: Plot of the energies of the first six levels for the
stretched potential~(\ref{eq:harmonic_potential_central}) for $n=4$. As $L$ increases, the levels with energy lower than $E_n$ form a ladder of two fold degenerate levels, while the states with higher energies asymptotically approach the free particle limit, i.e. plane waves.
Lower panel: plot of the associated wave-functions for $L/\xi=0.4$.}
\label{fig:v92_energy}
\end{figure}

Our next example assumes as seeding potential an harmonic one, i.e.
$V(x) =  \frac{m\omega^2}{2}x^2$, which admits eigenvalues $E_n= \hbar 
\omega (n+1/2)$ for $n\geq 0$ integer, with eigenfunctions
\begin{eqnarray} 
\psi_n(x) = \frac{( \tfrac{m\omega}{\pi \hbar} )^{1/4}}{\sqrt{2^n n!}} 
H_n\left(\sqrt{\frac{m\omega}{\hbar}} x \right) \exp[- \frac{m \omega x^2}{2\hbar}] \label{eigenfff}\;,  
 \end{eqnarray}  
 $H_n(x)$ being the $n$-th Hermite polynomial. 
Since for $n$ even, the energy wavefunction
admits a stationary point in $x=0$, we can stretched it by using 
 the potential 
\begin{eqnarray}
\label{eq:harmonic_potential_central}
V^{[L]}(x)=\left\{\begin{array}{cc}
\frac{m\omega^2}{2}(x+\frac{L}{2})^2&\mbox{$x\leq-\frac{L}{2}$}\\
&\\
E_n&\mbox{$|x|<\frac{L}{2}$}\\
&\\
\frac{m\omega^2}{2}(x-\frac{L}{2})^2&\mbox{$x\geq\frac{L}{2}.$}
\end{array}\right.
\end{eqnarray}
In order to determine the spectrum of the model, we can  resort in solving 
the associated Sturm-Liouville differential equation 
in each of the
three spatial domain separately, and then try to  match them with proper continuity conditions.
For this purpose we observe that the only solution
which that is square-integrable   for $x\leq L/2$ (resp. $x\geq -L/2$)
is given by the parabolic cylinder function~\cite{SOL1} 
$D_{\epsilon-\frac{1}{2}}(-\sqrt{2}\xi(x+\frac{L}{2}))$ 
(resp. $D_{\epsilon-\frac{1}{2}}(\sqrt{2}\xi(x+\frac{L}{2}))$), where for simplicity 
we set $\epsilon=\frac{E}{\hbar\omega}$ and $\xi=\sqrt{\frac{m\omega}{\hbar}}$ -- the latter
reducing to (\ref{eigenfff}) for $\epsilon$ semi-integer.
Accordingly introducing the rescaled quantities
$\epsilon_n=\frac{E_n}{\hbar\omega}$, $\gamma=\sqrt{\epsilon_n-\epsilon}$ and 
$\bar{k}=\sqrt{\epsilon-\epsilon_n}$, up to a normalization constant, 
the eigensolutions for  $E<E_n$  must have the form 
\begin{eqnarray}
\label{eq:harmonic_wavef_even}
\psi(x)=\left\{\begin{array}{cc}
\frac{\cosh(\sqrt{2}\xi\gamma\frac{L}{2})}{D_{\epsilon-\frac{1}{2}}(0)}D_{\epsilon-\frac{1}{2}}(-\sqrt{2}\xi(x+\frac{L}{2}))&\\
&\\
\cosh(\sqrt{2}\xi\gamma x)&\\
&\\
\frac{\cosh(\sqrt{2}\xi\gamma\frac{L}{2})}{D_{\epsilon-\frac{1}{2}}(0)}D_{\epsilon-\frac{1}{2}}(\sqrt{2}\xi(x-\frac{L}{2}))\;,&
\end{array}\right.
\end{eqnarray}
for even states
with quantization condition
\ba
\label{eq:harmonic_even_quant_cond}
\frac{D_{\epsilon+\frac{1}{2}}(0)}{D_{\epsilon-\frac{1}{2}}(0)}=-\gamma\tanh(\sqrt{2}\xi\gamma\frac{L}{2})\;, 
\ea
and
\begin{eqnarray}
\label{eq:harmonic_wavef_odd}
\psi(x)=\left\{\begin{array}{cc}
-\frac{\sinh(\sqrt{2}\xi\gamma\frac{L}{2})}{D_{\epsilon-\frac{1}{2}}(0)}D_{\epsilon-\frac{1}{2}}(-\sqrt{2}\xi(x+\frac{L}{2}))&\\
&\\
\sinh(\sqrt{2}\xi\gamma x)&\\
&\\
\frac{\sinh(\sqrt{2}\xi\gamma\frac{L}{2})}{D_{\epsilon-\frac{1}{2}}(0)}D_{\epsilon-\frac{1}{2}}(\sqrt{2}\xi(x-\frac{L}{2}))\;,&
\end{array}\right.
\end{eqnarray}
for the odd states, with quantization condition
\ba
\label{eq:harmonic_odd_quant_cond}
\frac{D_{\epsilon+\frac{1}{2}}(0)}{D_{\epsilon-\frac{1}{2}}(0)}=-\gamma\coth(\sqrt{2}\xi\gamma\frac{L}{2})\;. 
\ea
As $E>E_n$ we get instead:
\ba
\psi(x)=\left\{\begin{array}{cc}
\frac{\cos(\sqrt{2}\xi\bar{k}\frac{L}{2})}{D_{\epsilon-\frac{1}{2}}(0)}D_{\epsilon-\frac{1}{2}}(-\sqrt{2}\xi(x+\frac{L}{2}))&\\
&\\
\cos(\sqrt{2}\xi\bar{k}x)&\\
&\\
\frac{\cos(\sqrt{2}\xi\bar{k}\frac{L}{2})}{D_{\epsilon-\frac{1}{2}}(0)}D_{\epsilon-\frac{1}{2}}(+\sqrt{2}\xi(x-\frac{L}{2}))\;,&
\end{array}\right.
\ea
for even states and:
\ba
\psi(x)=\left\{\begin{array}{cc}
-\frac{\sin(\sqrt{2}\xi\bar{k}\frac{L}{2})}{D_{\epsilon-\frac{1}{2}}(0)}D_{\epsilon-\frac{1}{2}}(-\sqrt{2}\xi(x+\frac{L}{2}))&\\
&\\
\sin(\sqrt{2}\xi\bar{k}x)&\\
&\\
\frac{\sin(\sqrt{2}\xi\bar{k}\frac{L}{2})}{D_{\epsilon-\frac{1}{2}}(0)}D_{\epsilon-\frac{1}{2}}(+\sqrt{2}\xi(x-\frac{L}{2}))\;,&
\end{array}\right.
\ea
the associated  quantization conditions being respectively 
\ba
\label{eq:harmonic_even_quant_cond_up}
\frac{D_{\epsilon+\frac{1}{2}}(0)}{D_{\epsilon-\frac{1}{2}}(0)}&=&\bar{k}\tan(\sqrt{2}\xi\gamma\frac{L}{2})\;,\\
\label{eq:harmonic_odd_quant_cond_up}
\frac{D_{\epsilon+\frac{1}{2}}(0)}{D_{\epsilon-\frac{1}{2}}(0)}&=&-\bar{k}\cot(\sqrt{2}\xi\gamma\frac{L}{2})\;.
\ea
We start by noting that when $E=E_n$, then both Eq.~\eqref{eq:harmonic_even_quant_cond} and Eq.~\eqref{eq:harmonic_even_quant_cond_up} are satisfied, and substituting in Eq.~\eqref{eq:harmonic_wavef_even} 
we get the stretched counterpart of $\psi_n(x)$ as expected
(for this purpose it is useful to remind  that 
$D_n(x)=2^{-\frac{n}{2}}e^{-\frac{x^2}{4}}H_n(\frac{x}{\sqrt{2}})$). 
We also observe that  in the limit of large $L$, Eq.~\eqref{eq:harmonic_even_quant_cond} and Eq.~\eqref{eq:harmonic_odd_quant_cond} become identical,  resulting into  the formation of two fold degenerate state with energy below $E_n$ in closed analogy to what observed in
Sec.~\ref{sec:exc}. 
Finally  from Eq.~\eqref{eq:harmonic_even_quant_cond_up} and Eq.~\eqref{eq:harmonic_odd_quant_cond_up} we can verify that as $L$ becomes large the states with energy greater than $E_n$ tend to approach the plane waves limit and the energy
gaps tend to get compressed. 
This results are confirmed and summarized in the plot in Fig.~\ref{fig:v92_energy}, where the energy of the various levels is plotted as a function of $L$.

\section{\label{sec:conclusions}Conclusions}
The potentialities offered by the extension of matematerials from the conventional optical regime to electronic platforms where (at difference with photons) charged particles can strongly interact, is currently attracting a lot of attention, thus bringing  to the design of the so-called synthetic quantum metamaterials~\cite{Gabor}. From a theoretical point of view this naturally calls for the extension of the concept of stretched electromagnetic waves to stretched quantum states of massive particles, which is actually the goal of this manuscript.
Our construction hinges upon the one-to-one correspondence between the Helmhotz and the Schr\"odinger equations, leading to introduction of stretching potentials which admit spatially flat eigenfunctions in regions where the potential is constant as well. Notwithstanding the validity of such definition for arbitrary dimensions, herewith we mostly focused on the 1D scenario for which we have also provided a necessary and sufficient criterion: a stretching potential can be yielded only by properly deforming a seeding potential, characterized by explicitly non-stretched eigenstates, by inserting in it flat regions bearing exactly the same energy of the selected eigenfunction. 
This can find applications in nowadays quantum technologies, such as the electron-potential engineering of nanowires~\cite{Bryllert,Svensson,Roddaro,Prete}, or the implementation of nano-structures whose working principle is based on coupling the nuclear spins to electron transport~\cite{chesi}. An even reacher plethora of applications is offered by 2D geometries, such as the one commonly found in semiconductor electronic waveguides. To this end, we showed that by slowly varying the length of the flat region of the potential along one coordinate (say $x$)  as a function of the other coordinate (say $y$), our scheme can be easily extended to non-trivial two dimensional configurations.  This observation brings a straightforward connection between our scheme and the adiabatic theorem, thus paving the way to a feasible practical implementation of our proposal also in 1D setups. In fact, a given stretching potential can be easily generated by properly tuning a set of time-dependent control parameters labelling the associated seeding potential, so as to transform the selected eigenstate into a the target stretch-quantum state.

\acknowledgements 
The Authors would like to thank Fabio Taddei for useful comments.

\appendix
\section{Justifying the ansatz of Eq.~(\ref{EASY})} 
\label{APPA} 
As anticipated in the Sec.~\ref{sec:high}, under slow varying assumptions of the
function $L(y)$, Eq.~(\ref{EASY}) arguably provides a good  approximation for a solution 
of Eq.~(\ref{defimpo}) when we enforce the stretching 
$V(x) \rightarrow  V^{[{L}(y)]}(x)$ of the confining potential. 
 A simple way to see this
is to notice that by adopting the ansatz~(\ref{EASY}), the extra contribution 
(\ref{dde})  becomes
\begin{eqnarray} 
&&\Delta(x,y) \Big|_{\psi_0^{[L(y)]}(x)} 
 = - \Theta(x'- \bar{x}) \Big[
  \partial_{x'} \psi_0(x')    {L}''(y)   \\ \nonumber 
 &&  + \Big(2i k \partial_{x'} \psi_0(x') + \partial^2_{x'} \psi_0(x')  {L}'(y)\Big) {L}'(y)
  \Big]_{x'=x-{L}(y)}  \;,  \label{QUESTO} 
\end{eqnarray} 
which gets suppressed in the limit where ${L}(y)$ is almost constant (here $\Theta(x)$ stands for the
Heaviside step function). Specifically, for assigned $y$, we notice that 
$\ell^{(2)}$ norm of $\Delta(x,y)$  under the ansatz is given by 
\begin{widetext} 
\begin{eqnarray} 
\left\|\Delta(x,y)\right\|^2 &=& \int^{\infty}_{-\infty}  dx  |\Delta(x,y) |^2  
= \int_{\bar{x}}^\infty dx \Big|
  \partial_{x} \psi_0(x)   ( {L}''(y)  + 2i k {L}'(y)) 
  +  \partial^2_{x} \psi_0(x)  \big({L}'(y) \big)^2
  \Big|^2 \nonumber  \\ &\leq& 
  \int_{-\infty}^\infty dx \Big|
  \partial_{x} \psi_0(x)   ( {L}''(y)  + 2i k {L}'(y)) 
  +  \partial^2_{x} \psi_0(x)  \big({L}'(y) \big)^2
  \Big|^2 \nonumber \\
     &\leq & | {L}''(y)  + 2i k {L}'(y)|^2 \int_{-\infty}^\infty dx | \partial_{x} \psi_0(x) |^2
  +  | {L}'(y)|^4 \int_{-\infty}^\infty dx | \partial^2_{x} \psi_0(x) |^2
  \nonumber \\
    &&+ 2\big|{L}'(y) \big|^2  | {L}''(y)  + 2i k {L}'(y)) |
  \sqrt{ \int_{-\infty}^\infty dx | \partial_{x} \psi_0(x) |^2 \int_{-\infty}^\infty dx | \partial^2_{x} \psi_0(x) |^2} 
 \nonumber \\
 &=& \left(
  | {L}''(y)  + 2i k {L}'(y)| \sqrt{  \; \int_{-\infty}^\infty dx  |\partial_{x} \psi_0(x)|^2 }
     +    \big|{L}'(y) \big|^2\sqrt{ \int_{-\infty}^\infty dx | \partial^2_{x} \psi_0(x) |^2}    
  \right)^2 \nonumber\\
   &=&  \left(
  | {L}''(y)  + 2i k {L}'(y)| \sqrt{ \langle  \hat{p}_x^2\rangle_0   }/\hbar 
     +    \big|{L}'(y) \big|^2\sqrt{ \langle  \hat{p}_x^4\rangle_0 }/\hbar^2     
  \right)^2\;, 
\end{eqnarray} 
\end{widetext} 
where $\langle \cdots \rangle_0$ represents the expectation value with respect to
$\psi_0(x)$, where $\hat{p}^2_x$ indicates the transverse kinetic energy (in deriving the above expression we invoked
the Chaucy-Swartz inequality), and where hereafter  $L'(x):= \partial_y L(y)$ and $L''(x):= \partial^2_y L(y)$.
On the contrary for the $\ell^2$ norm of the first contribution in the r.h.s. of Eq.~(\ref{eq:standard_schroedinger2Dnew})
i.e. the quantity
$f(x,y):=-\frac{2m}{\hbar^2}\big[ E-V^{[{L}(y)]}(x) \big] {\psi}_{0}^{[L(y)]}(x)$,
we get 
\begin{eqnarray} 
\left\|f(x,y)\right\|^2 &=& \int^{\infty}_{-\infty}  dx  |f(x,y) |^2 =
 \int^{\infty}_{-\infty}  dx  |\partial^2_x{\psi}_{0}^{[L(y)]}(x) |^2 \nonumber \\
 &=&  \int_{-\infty}^{\bar{x}}  dx  |\partial^2_x{\psi}_{0}(x) |^2+
  \int^{\infty}_{\bar{x}+L}  dx  |\partial^2_x{\psi}_{0}(x-L) |^2 \nonumber \\
  &=&  \int_{-\infty}^{\infty}  dx  |\partial^2_x{\psi}_{0}(x) |^2 =  \langle  \hat{p}_x^4\rangle_0 /\hbar^4\;,
\end{eqnarray} 
where in the first line we used the fact that since ${\psi}_{0}^{[L(y)]}(x)$ is an explicit eigenfunction of the stretched potential one has
$f(x,y)=
\partial^2_x{\psi}_{0}^{[L(y)]}(x)$.
Putting all this together we can conclude that
as long as the following inequality holds for all $y$, 
\begin{eqnarray} 
{\frac{ \langle  \hat{p}_x^4\rangle_0}{\langle  \hat{p}_x^2\rangle_0 \hbar^2} }
\gg \frac{{|L''(y)|^2+ 4 k^2 |L'(y)|^2}}{(1- \left |L'(y)\right|^2)^2}\;,
\end{eqnarray} 
we can ensure that under the ansatz (\ref{EASY})  one has 
\begin{eqnarray} 
\left\|f(x,y)\right\| \gg \left\|\Delta(x,y)\right\|\;,
\end{eqnarray}
meaning that the second contribution on the r.h.s. of Eq.~(\ref{eq:standard_schroedinger2Dnew}) is much smaller than  the first. 
Accordingly, when integrating  such differential equation over a not
too large integration interval
 we can neglect 
the contribution of $\Delta(x,y)$, hence justifying the approximation~(\ref{EASY}).

\end{document}